\documentclass[twocolumn,prl,amsmath,amssymb,superscriptaddress]{revtex4-1}
\usepackage[utf8]{inputenc}
\usepackage{amsfonts,amsmath,amssymb}
\usepackage{graphicx}
\usepackage{bm} 
\usepackage{mathrsfs}
\usepackage{subfigure}
\usepackage{textcomp}
\usepackage{microtype}
\usepackage{hyperref}
\usepackage[flushleft]{threeparttable}
\usepackage[per-mode=symbol]{siunitx}
\usepackage[version=3]{mhchem}

\usepackage{booktabs}
\DeclareSIUnit\intensity{\watt\per\centi\meter\squared}
\DeclareSIUnit\fieldstrength{\volt\per\centi\meter}
\usepackage{xspace}

\DeclareUnicodeCharacter{2009}{\,}

\newcommand{\degree}{\ensuremath{^\circ}}%

\newlength{\figwidth}
\newlength{\figwidthwide}
\let\orgautoref\autoref
\providecommand{\Autoref}{%
  \def\equationautorefname{Equation}%
  \def\figureautorefname{Figure}%
  \def\subfigureautorefname{Figure}%
  \def\tableautorefname{Table}
  \orgautoref}
\renewcommand{\autoref}{%
  \def\equationautorefname{Eq.}%
  \def\figureautorefname{Fig.}%
  \def\subfigureautorefname{Fig.}%
  \orgautoref}

\usepackage{color}
\usepackage[normalem]{ulem}
\definecolor{darkgreen}{rgb}{0.0,0.7,0.0}

\begin{document}

\title{Quantum-State-Sensitive Detection of Alkali Dimers on Helium Nanodroplets by Laser-Induced Coulomb Explosion}

\author{Henrik H. Kristensen}
\affiliation{Department of Physics and Astronomy, Aarhus University, Ny Munkegade 120, DK-8000 Aarhus C, Denmark}
\author{Lorenz Kranabetter}
\affiliation{Department of Chemistry, Aarhus University, Langelandsgade 140, DK-8000 Aarhus C, Denmark}
\author{Constant A. Schouder}
\affiliation{Department of Chemistry, Aarhus University, Langelandsgade 140, DK-8000 Aarhus C, Denmark}
\author{Christoph Stapper}
\affiliation{Faculty of Chemistry and Pharmacy, University of Würzburg, Am Hubland, Campus Süd, D-97074 Würzburg, Germany}
\author{Jacqueline Arlt}
\affiliation{Department of Chemistry, Aarhus University, Langelandsgade 140, DK-8000 Aarhus C, Denmark}
\author{Marcel Mudrich}
\affiliation{Department of Physics and Astronomy, Aarhus University, Ny Munkegade 120, DK-8000 Aarhus C, Denmark}
\author{Henrik Stapelfeldt}
\email[]{henriks@chem.au.dk}
\affiliation{Department of Chemistry, Aarhus University, Langelandsgade 140, DK-8000 Aarhus C, Denmark}

\date{\today}

\begin{abstract}
Rubidium dimers residing on the surface of He nanodroplets are doubly ionized by an intense fs laser pulse leading to fragmentation into a pair of \ce{Rb^+} ions. We show that the kinetic energy of the \ce{Rb^+} fragment ions can be used to identify dimers formed in either the X $^1\Sigma_{\mathrm{g}}^+$ ground state or in the lowest-lying triplet state, a $^3\Sigma_{\mathrm{u}}^+$. From the experiment, we estimate the abundance ratio of dimers in the a and X states as a function of the mean droplet size and find values between 4:1 and 5:1. Our technique applies generally to dimers and trimers of alkali atoms, here also demonstrated for \ce{Li2}, \ce{Na2}, and \ce{K2}, and will enable fs time-resolved measurements of their rotational and vibrational dynamics, possibly with atomic structural resolution.
\end{abstract}

\maketitle

For more than 25 years helium nanodroplets have served as an exciting medium for exploring the structure and dynamics of impurities that can be either embedded inside the droplets or deposited on their surface~\cite{toennies_superfluid_2004,choi_infrared_2006,mauracher_cold_2018,chatterley_long-lasting_2019,thaler_long-lived_2020}. Most impurities, which encompass atoms, molecules, and clusters or complexes hereof, are located in the interior of the droplets~\cite{toennies_superfluid_2004,barranco_helium_2006}. Exceptions are alkali atoms, small clusters of alkali atoms, and partly alkaline earth atoms. These systems interact very weakly with helium and, as a result, they reside on the surface of the droplets~\cite{ancilotto_sodium_1995,stienkemeier_electronic_2001,toennies_superfluid_2004,barranco_helium_2006,barranco_helium_2006, ren_surface_2007,stark_critical_2010}.

Alkali dimers~\cite{stienkemeier_laser_1995,bruhl_triplet_2001,aubock_triplet_2007,ernst_cesium_2006,allard_investigation_2006} and trimers~\cite{stienkemeier_laser_1995,nagl_heteronuclear_2008,nagl_high-spin_2008} on He droplets have been the subject of many studies since the first experimental report in 1995~\cite{stienkemeier_laser_1995}. Experimental investigations relied mainly on vibrationally resolved spectroscopy of electronically excited states induced by absorption of visible or ultraviolet light. The most common methods applied have been laser-induced fluorescence and resonance-enhanced multiphoton ionization using cw, ns and fs lasers~\cite{mudrich_photoionisaton_2014}. These techniques established that the clusters are localized on the surface of the droplets, and that alkali dimers and trimers preferentially form in the highest spin states, i.e., triplet dimers and quartet trimers~\cite{stienkemeier_laser_1995,higgins_helium_1998,bruhl_triplet_2001,mudrich_formation_2004}. This contrasts the situation encountered with conventional gas-phase sources where mainly low-spin state dimers and trimers are present.

When an alkali dimer forms, its binding energy is dissipated by the helium droplet through evaporation of helium atoms which may cause desorption of the dimer from the droplet surface. Since the amount of internal energy released when forming the dimer in the singlet ground state greatly exceeds that released when forming the lowest triplet state, the latter has a higher chance to remain attached to the droplet. This leads to a dominance of droplets with triplet dimers~\cite{higgins_helium_1998,bunermann_modeling_2011}; yet both spectroscopy and dynamics of dimers in their singlet state have still been possible~\cite{stienkemeier_laser_1995,claas_wave_2006,ernst_cesium_2006,mudrich_spectroscopy_2009,laforge_highly_2019,ltaief_electron_2020}. Surprisingly, only a single work so far reported the abundance ratio of triplet to singlet dimers on helium nanodroplets~\cite{bunermann_modeling_2011}. An experimental method to quantitatively characterize this ratio, and which experimental parameters it depends on, has, to our knowledge, not been demonstrated.

In the present work, we introduce Coulomb explosion, induced by an intense fs laser pulse, as a means of studying alkali dimers on He droplets. Notably, we show that Coulomb explosion allows determination, within a single measurement, of whether dimers are formed in the X or in the a state. Our results open unexplored opportunities for measuring fs time-resolved vibrational and rotational motion of alkali dimers on a surface and for determining how the droplet-dimer coupling influences these motions.

\begin{figure}
\includegraphics[width=8.5 cm]{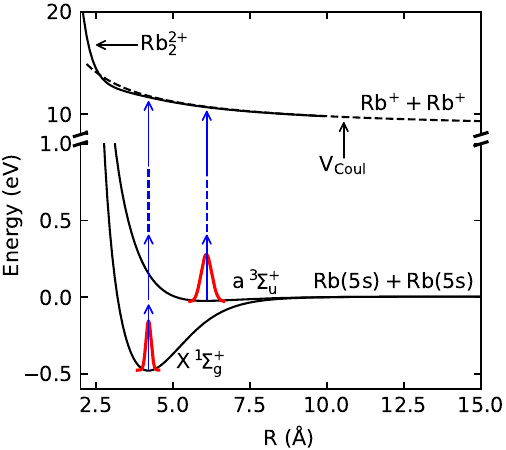}%
\caption{Energy diagram of the \ce{Rb} dimer showing the potential curves for the X $^1\Sigma_{\mathrm{g}}^+$ and a $^3\Sigma_{\mathrm{u}}^+$ states~\cite{allouche_transition_2012} and for \ce{Rb2^{2+}}~\cite{frank_jensen}. The dashed curve depicts the Coulomb potential. The red shapes show the square of the wave function of the vibrational ground state in the X and a potentials. The blue vertical arrows represent the laser photons (not to scale) to illustrate the double ionization process that leads to Coulomb explosion into a pair of \ce{Rb+} ions.}
\label{fig:Rb2_curves}
\end{figure}

The principle of our method is to first doubly ionize the alkali dimers by multiphoton absorption from an intense fs laser pulse as illustrated in \autoref{fig:Rb2_curves} for \ce{Rb2}. Hereby, the vibrational wave function of \ce{Rb2} is projected onto the potential curve of \ce{Rb2^{2+}}. Only one such curve exists, due to the closed-shell electron configuration of \ce{Rb2^{2+}}. For $R \geq$ 3.0  {\AA}, it is to a good approximation given by the repulsive Coulomb potential, $V_\text{Coul}(R)$ = 14.4~eV/$R$[{\AA}], where $R$ is the internuclear distance.  The Rb dimers are expected to share the 0.37~K temperature of the droplets~\cite{hartmann_rotationally_1995} and therefore solely be populated in the vibrational ground state. The corresponding equilibrium distance, $R_\text{eq}$, of the X and the a state, is 4.21 and 6.06~{\AA}, respectively, see \autoref{fig:Rb2_curves}~\cite{allouche_transition_2012}. After double ionization, the dimers in the X (a) state therefore acquire a potential energy  $V_\text{Coul}(R_\text{eq})$~=~3.42 eV (2.38 eV). The internal electrostatic repulsion of \ce{Rb2^{2+}} results in Coulomb explosion and conversion of $V_\text{Coul}(R_\text{eq})$ into kinetic energy $E_\text{kin}$ of the \ce{Rb+} ion fragments, i.e., each \ce{Rb+} ion acquires a final $E_\text{kin}$ of $\frac{1}{2}V_\text{Coul}(R_\text{eq})$.  Experimentally, we measured $E_\text{kin}$  and observed two distinct peaks centered at values close to $\frac{1}{2}V_\text{Coul}(R_\text{eq})$ for the X and a states. Thus, these two peaks enable an unambiguous identification of the initial electronic quantum state of the Rb dimers.

\begin{figure}
\includegraphics[width=8.5 cm]{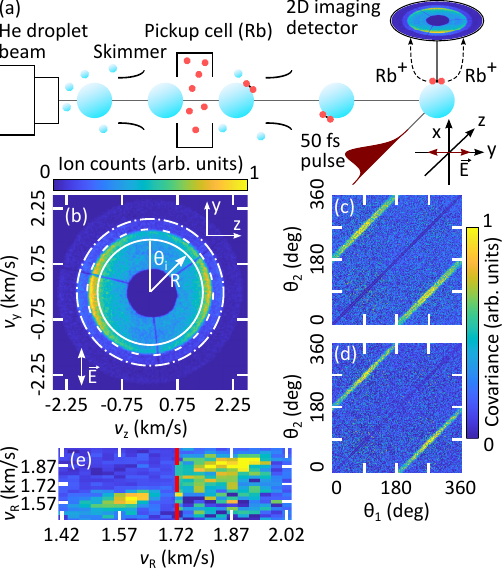}%
\caption{(a) Schematic of the key elements in the experiment. (b) 2D-velocity image of $^{85}$Rb$^+$ ions for $T_\text{nozzle}$~=~12~K. The annotated white circles mark the regions of the inner and outer channel, see text. The polarization direction of the laser pulse is displayed in the bottom left corner. (c), (d) Angular covariance maps for the inner and the outer channel. (e) Radial covariance map. The region after the red dashed line (1.72--2.02~km/s) has been saturated to highlight the presence of the outer channel. The color bar to the right applies to (c)--(e).}
\label{fig:setup_data}
\end{figure}

\begin{figure}
\includegraphics[width=8.5 cm]{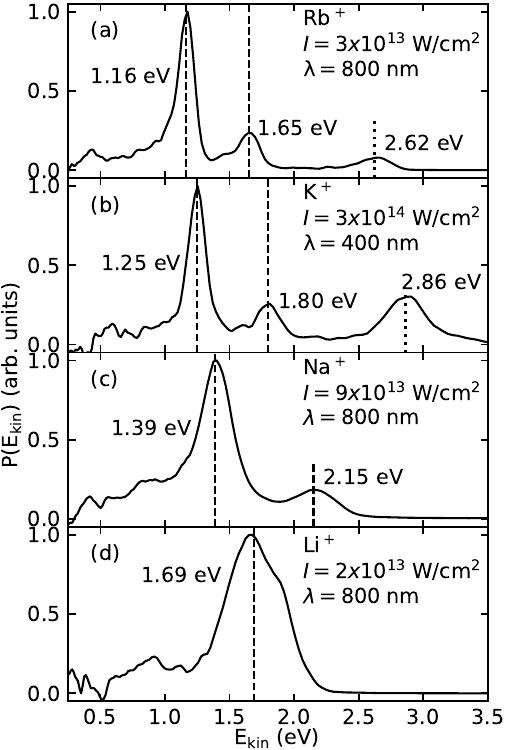}%
\caption{(a) Kinetic energy spectrum of \ce{Rb+} ions extracted from the 2D velocity image in \autoref{fig:setup_data}(b). (b)--(d) Kinetic energy spectrum of \ce{K+}, \ce{Na+} and \ce{Li+} ions, also extracted from corresponding 2D velocity images (not shown). $T_\text{nozzle}$ was 12~K for the data in (a) and 11~K for the data in (b)--(d). The vertical dashed (dotted) lines mark the center of the peaks ascribed to dimers (trimers).}
\label{fig:Rb2_energies}
\end{figure}

The experimental setup is similar to that described previously and only a few details are given here~\cite{shepperson_strongly_2017}. A continuous helium droplet beam is created by expanding high purity helium at 25 bar into vacuum through a 5~$\mu$m diameter nozzle cooled to $T_\text{nozzle}$ = 11--16~K. The droplets travel through a pickup cell containing a gas of Rb atoms, see \autoref{fig:setup_data}(a). The Rb vapor pressure is adjusted such that some of the droplets pick up two Rb atoms, which leads to formation of Rb dimers on the droplet surfaces~\cite{stienkemeier_use_1995, bruhl_triplet_2001, trimer_note}. Further downstream, inside a velocity map imaging (VMI) spectrometer, the doped droplet beam is crossed by a pulsed linearly polarized, focused laser beam. The duration of the pulses is 50 fs and their central wavelength $\lambda$ and intensity $I$ are given in \autoref{fig:Rb2_energies}. The Rb$^+$ ions created by the laser pulses are projected by the VMI spectrometer onto a 2D imaging detector backed by a CCD camera. The detector is gated such that only $^{85}$Rb$^+$ ions are detected. Frames, containing the ion hits from 10 laser pulses, are collected from the camera.

\Autoref{fig:setup_data}(b) shows a 2D velocity image of \ce{Rb+} ions, obtained by stacking 20,000 frames. The absence of signal in the center and in three radial stripes comes from a metal disk~\cite{schouder_laser-induced_2020, chatterley_laser-induced_2020} and its supports. This was installed in front of the detector to block the \ce{Rb+} ions originating from ionization of Rb atoms on droplets that picked up only one Rb atom or ionization of isolated Rb atoms that diffused into the VMI spectrometer. Outside of the blocked center, corresponding to ions with higher kinetic energy, two channels are observed. The first, termed the inner channel, consists of the ions between the annotated full and dashed circles and is centered at a radius of 1.6~km/s. The second, termed the outer channel, consists of the ions between the annotated dashed and dot-dashed circles. It is centered at 1.9~km/s. In addition, there is a weak channel centered at around 2.3~km/s.

To understand the origin of the channels, we determine the kinetic energy distribution, $P(E_\text{kin})$, of the \ce{Rb+} ions. This is done by first Abel inverting the ion image~\cite{roberts_toward_2009} to retrieve the radial velocity distribution, and then converting to $P(E_\text{kin})$ by applying an energy calibration for the VMI spectrometer and the standard Jacobian transformation~\cite{schouder_laser-induced_2020}. \Autoref{fig:Rb2_energies}(a) shows $P(E_\text{kin})$. It contains two distinct peaks with central positions at 1.16 and 1.65~eV (found by fits with a Gaussian function) as marked by the vertical dashed lines. These values are very close to the final $E_\text{kin}$ each \ce{Rb+} ion would acquire if \ce{^85Rb2^{2+}} breaks apart starting from $R_\text{eq}$ for the a and X state, i.e. 1.19 and 1.71 eV~\footnote{57 (43) \% of the \ce{^85Rb} ions detected come from \ce{^85Rb2} (\ce{^85Rb^87Rb}). $E_\text{kin}$ of the ions from \ce{^85Rb^87Rb} is 15 meV higher than ions from \ce{^85Rb2}}. Therefore, we interpret the peaks at 1.16 and 1.65 eV as ions stemming from Coulomb explosion of \ce{Rb2} in the a and in the X state, respectively. The match of the peak positions and $E_\text{kin}$ expected for Coulomb explosion of \ce{Rb2} shows that the interaction between the recoiling \ce{Rb+} ions and the He droplet only marginally changes their kinetic energy as the Coulomb explosion proceeds.

To test our interpretation, we explore if correlations exist between the emission directions of the \ce{Rb+} ions originating from the same laser pulse because the Coulomb explosion should produce pairs of \ce{Rb+} ions departing back to back. It is done by a statistical analysis, termed angular covariance maps~\cite{hansen_control_2012,frasinski_covariance_2016,vallance_covariance-map_2021}, for both the inner and outer channel. The results, displayed in \autoref{fig:setup_data}(c)--(d), show two distinct diagonal lines, centered at $\theta_2 = \theta_1 + \SI{180}{\degree}$ and $\theta_2 = \theta_1 - \SI{180}{\degree}$, in each channel, where $\theta_i$, $i$=1,2 is the angle between an ion hit and the vertical center line, \autoref{fig:setup_data}(b). These lines demonstrate a $\SI{180}{\degree}$ correlation in the \ce{Rb+} ion emission directions and thereby corroborate that Coulomb explosion is the origin of the ions, analogous to previous works~\cite{shepperson_strongly_2017,pickering_alignment_2018,schouder_structure_2019}.

The transverse width of the diagonal lines is \SI{19}{\degree} for the inner and \SI{14}{\degree} for the outer channel. In comparison, for a gas phase molecule, like \ce{I2}, the width is $\sim$~\SI{1}{\degree}~\cite{shepperson_strongly_2017}. Thus, the observed diagonal lines indicate that the He droplet slightly distorts the direction of the recoiling \ce{Rb+} ions. We note that the ion yield is higher at \SI{90}{\degree} and \SI{270}{\degree} than at \SI{0}{\degree} and \SI{180}{\degree}, as seen in both the ion image and in the angular covariance maps. We ascribe this to resonant enhancement of the multiphoton absorption process due to electronically excited states of \ce{Rb2}~\cite{sieg_desorption_2016}. This picture is supported by observations of different angular anisotropies for other wavelengths of the laser pulse and for other alkali dimers.

We also determined the radial covariance map, i.e., the covariance map for the radial distributions of the upper and lower half on the ion image in \autoref{fig:setup_data}(a)~\cite{christiansen_laser-induced_2016}. The two islands in the radial covariance map, displayed in \autoref{fig:setup_data}(e), show correlations between two ions each with a velocity $v_R$ around 1.6~km/s or around 1.9~km/s. Such correlations are expected for Coulomb explosion of \ce{Rb2} in either the a or in the X state because the two \ce{Rb+} fragments obtain the same final velocity. All in all, the information from $P(E_\text{kin})$ and from the angular and radial covariance maps allows us to unambiguously identify the \ce{Rb+} ions in the inner channel and outer channels, or equivalently in the 1.16 and 1.65 eV peaks, as originating from Coulomb explosion of Rb dimers in the a and X state, respectively. Concerning the ions in the small peak at 2.62 eV, we found that they originate from Coulomb explosion of Rb trimers with an equilateral shape into three \ce{Rb+} ions~\cite{kranabetter_2021}.

\begin{figure}
\includegraphics[width=8.5 cm]{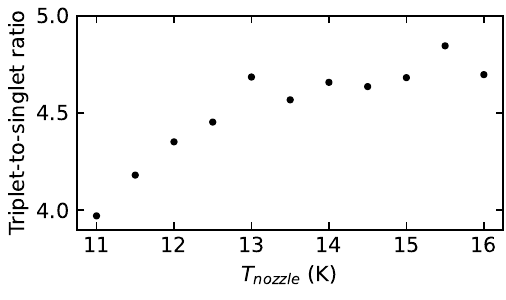}%
\caption{Ratios of Rb$^+$ fragments originating from Coulomb exploded triplet and singlet Rb$_2$ for different nozzle temperatures.}
\label{fig:ratios}
\end{figure}

From the ratio between the integrated area of the 1.16 and the 1.65 eV peaks in \autoref{fig:Rb2_energies}(a), we extract a measure for the ratio of the number of \ce{Rb} dimers populated in the a state and in the X state, hereafter termed the triplet-to-singlet ratio. We determined this ratio for a series of different nozzle temperatures between 11 and 16~K corresponding to an average number of He atoms in the droplets of $\sim$~4000 (15000) for $T_\text{nozzle}$=16~K (11~K)~\cite{toennies_superfluid_2004}. The results, depicted in \autoref{fig:ratios}, show that the triplet-to-singlet ratio increases when $T_\text{nozzle}$ increases, i.e., when the droplet size decreases. This is the expected behavior because as the dimers form on the surface, $\sim$ 800 (50) He atoms must evaporate from the droplet to remove the 0.48 eV (30 meV) binding energy of the X (a) state, see \autoref{fig:Rb2_curves}~\cite{theisen_rb_2011}. For the X state, the number of evaporated He atoms is so large that the surface gets distorted to an extent that the dimer may desorb. The significance of the distortion increases when the droplets become smaller and thus the probability of finding \ce{Rb2} in the X state on the droplet surface decreases.

\Autoref{fig:ratios} shows that the triplet-to-singlet ratio of \ce{Rb2} varies between 4.0:1 and 4.8:1. To our knowledge, the only previous report on the triplet-to-singlet ratio was for \ce{Na2}~\cite{bunermann_modeling_2011}. Based on combined experimental and simulated studies, this work reported values of 50:1 for droplets containing 5000--20000 He atoms. To critically assess how accurately our method determines the absolute value of the triplet-to-singlet ratio, we remark that the energy needed to doubly ionize the X state is 12.1 eV but only 10.7 eV for the a state, see \autoref{fig:Rb2_curves}. For multiphoton ionization, this energy difference could give a higher double ionization probability for \ce{Rb2} in the a-state if $I$ is too low. Consequently, we measured the triplet-to-singlet ratio for a range of different laser intensities and found that it saturated at a value independent of $I$ when $I$ $>$ 1.0$\times$10$^{13}$~W/cm$^2$. The data presented in \autoref{fig:Rb2_energies}(a) are recorded in the saturation regime.

Another factor that can influence if the measured triplet-to-singlet ratio reflects the true triplet-to-singlet ratio is single ionization of the dimers because, in principle, the branching ratio of single and double ionization could be different for the a and the X state. Experimentally we cannot directly decide if such a difference exists due to the metal plate in front of the detector. However, the use of a high intensity laser pulse ensures that most \ce{Rb} dimers are doubly ionized rather than singly ionized. As such, we believe that our method provides a fairly accurate measure for the triplet-to-singlet ratio. For the largest droplets obtained at $T_\text{nozzle}$=11~K, the measured ratio of 4.0:1 approaches the 3:1 ratio expected if only the spin statistical factors determine the abundance of the two states~\cite{higgins_helium_1998, bunermann_modeling_2011}.

To demonstrate the applicability of Coulomb explosion for identifying the a and X state of other alkali dimers, we also conducted experiments on \ce{K2}, \ce{Na2,} and \ce{Li2}. The resulting kinetic energy spectra are shown in \autoref{fig:Rb2_energies}(b)--(d). As for \ce{Rb2}, we find that the position of the peaks marked by the vertical dashed lines closely match $\frac{1}{2}V_\text{Coul}(R_\text{eq})$ for the a and X states. This allows us to identify the origin of the peaks as summarized in~\autoref{tab:energies}. Similar to the \ce{Rb2} measurement, angular and radial covariance maps of the \ce{K+}, \ce{Na+}, and \ce{Li+} ions corroborate the assignment of the peaks in $P(E_\text{kin})$ to Coulomb explosion of dimers in the a and X states.

\begin{table}
\centering
\begin{tabular}{ccccc}
\toprule
      & \multicolumn{2}{l}{Peaks in P($E_\text{kin}$) (eV)} & \multicolumn{2}{l}{$\frac{1}{2} E_{Coul}(R_\text{eq}$) (eV)} \\ \midrule
      & Inner                  & Outer                 & a $^3\Sigma_u^+$         & X $^1\Sigma_g^+$         \\ \midrule
\ce{Rb_2} & 1.16                   & 1.65                   & 1.19                     & 1.71                     \\
\ce{K_2}  & 1.25                   & 1.80                   & 1.26                     & 1.83                     \\
\ce{Na_2} & 1.39                   & 2.15                   & 1.39                     & 2.34                     \\
\ce{Li_2} & 1.69                   & -                      & 1.73                     & 2.69					  \\
\bottomrule
\end{tabular}
\caption{$E_\text{kin}$ of the fragment ions from Coulomb exploded alkali dimers. The $R_\text{eq}$ values are taken from~\cite{deiglmayr_calculations_2008, allouche_transition_2012, bauer_accurate_2019}.}
\label{tab:energies}
\end{table}

For \ce{K2} the triplet-to-singlet ratio is 3.7:1 and for \ce{Na2} it is 6.3:1. Unlike for \ce{Rb2}, we did not systematically perform measurements at many different laser intensities but the intensities used were so high that the measurements were most likely performed in the saturation regime, in particular for \ce{K2}. In the case of \ce{Li2}, only the triplet state is observed consistent with previous works~\cite{higgins_helium_1998, lackner_spectroscopy_2013}. Note that for \ce{K}, the peak centered at 2.86 eV can be assigned as originating from Coulomb explosion of the \ce{K} trimer. For \ce{Na}, the trimer peak is also observed at higher vapor pressures in the pickup cell~\cite{kranabetter_2021}.

In summary, we introduced Coulomb explosion as a method to identify, within a single measurement, whether alkali dimers, \ce{Ak2}, on the surface of He nanodroplets are formed in the X or in the a state. The identification relied on the measurement of the kinetic energies of the \ce{Ak+} fragment ions, which are significantly different for dimers in the a and X state due to the difference in $R_\text{eq}$. We point to two opportunities based on our results. The experimental $P(E_\text{kin})$ may be used not just to identify the two quantum states but also to measure the vibrational wave function. The alkali dimers are ideal systems for wave function imaging~\cite{zeller_imaging_2016,schouder_laser-induced_2020} through Coulomb explosion because of the one-to-one correspondence between $E_\text{kin}$ and $R$ due to the single dicationic state. Particularly interesting would be to image the time-dependent evolution of vibrational wave packets created through stimulated Raman transitions by a fs pump pulse~\cite{claas_wave_2006,shu_femtochemistry_2017}, and to observe if vibrational relaxation from coupling to the droplet distorts the wave packets. Also, it should be possible to use the emission direction of the \ce{Ak+} fragment ions to explore if laser-induced alignment of alkali dimers on a surface can be realized. Ongoing experiments in our laboratory show that this is indeed possible and that time-resolved rotational dynamics, induced by a fs alignment pulse, enables determination of the, hitherto unknown, rotational structure of alkali dimers on the He droplet surface.

\begin{acknowledgments}
We acknowledge the helpful discussions with Frank Stienkemeier and Frank Jensen. H.S. acknowledges support from Villum Fonden through a Villum Investigator Grant No. 25886.
\end{acknowledgments}


\begin{thebibliography}{49}
\bibitem [{tri()}]{trimer_note}%
  \BibitemOpen
  \href@noop {} {\bibinfo  {journal} {\noindent The observation of a weak
  signal from Rb trimers, see discussion of Fig. 3, shows that a very small
  fraction of the droplets even pick up three Rb atoms}\ }\BibitemShut
  {NoStop}%
\end{thebibliography}
\end{document}